\documentclass[%
 reprint,
showpacs,
 amsmath,amssymb,
 aps,
]{revtex4-1}

\usepackage{txfonts,graphicx}
\usepackage{dcolumn}
\usepackage{bm}

\begin{document}

\title{Space-Time Metamaterials.}

\author{Andrei Rogov and Evgenii Narimanov}
\affiliation{School of Electrical and Computer Engineering, and Birck Nanotechnology Center, Purdue University, West Lafayette IN 47907}

\date{\today}

\begin{abstract}
Despite more than a decade of active research, the fundamental problem of material loss 
remains a major obstacle in  fulfilling the promise of the recently emerged fields of metamaterials 
and plasmonics to bring in revolutionary practical applications. In the present work, we demonstrate that the
problem of strong material absorption that is inherent to plasmonic systems  and metamaterials based 
on plasmonic components, can be addressed by utilizing the time dimension.  By matching the pulse profile to the actual response of a lossy metamaterial, this  approach
allows to offset the effect of the material absorption. The existence of the corresponding solution relies on the fundamental property of causality, that relates the absorption in the medium to the variations in the frequency-dependent time delay introduced by the 
material, via the Kramers-Kronig relations. We demonstrate that the proposed space-time approach can be applied 
to a broad range of metamaterial-based and plasmonic systems, from hyperbolic media to metal optics and 
new plasmonic materials.
\end{abstract}

\maketitle

 \section{Introduction}

The fields of metamaterials and plasmonics promise a broad range of exciting applications -- from
electromagnetic invisibility and cloaking \cite{cloaking} to negative refraction and super-resolution imaging \cite{pendry2000} to subwavelength field localization, confinement and  amplification
in plasmonic resonators. \cite{Vlad-book,spacer} However, the performance of practical plasmonics- and metamaterials- based applications is severely limited by losses \cite{Vlad-book} -- e.g. it is the material absorption that ``confines'' the superlens to the 
near-field operation. \cite{OL-SL} Despite multiple attempts to remove this stumbling block with new materials \cite{Sasha_AM}  or incorporating material gain in the composite design \cite{Vlad_Science}, the  (nearly) ``lossless metal'' \cite{Jacob_APL} that would allow the evanescent field control and amplification promised by metamaterial research for nearly two decades since the seminal work of J. Pendry, \cite{pendry2000}  remains an elusive goal.\cite{Sasha_AM}

\begin{figure*}[htbp]
  \centering
  \includegraphics[width=0.9\textwidth]{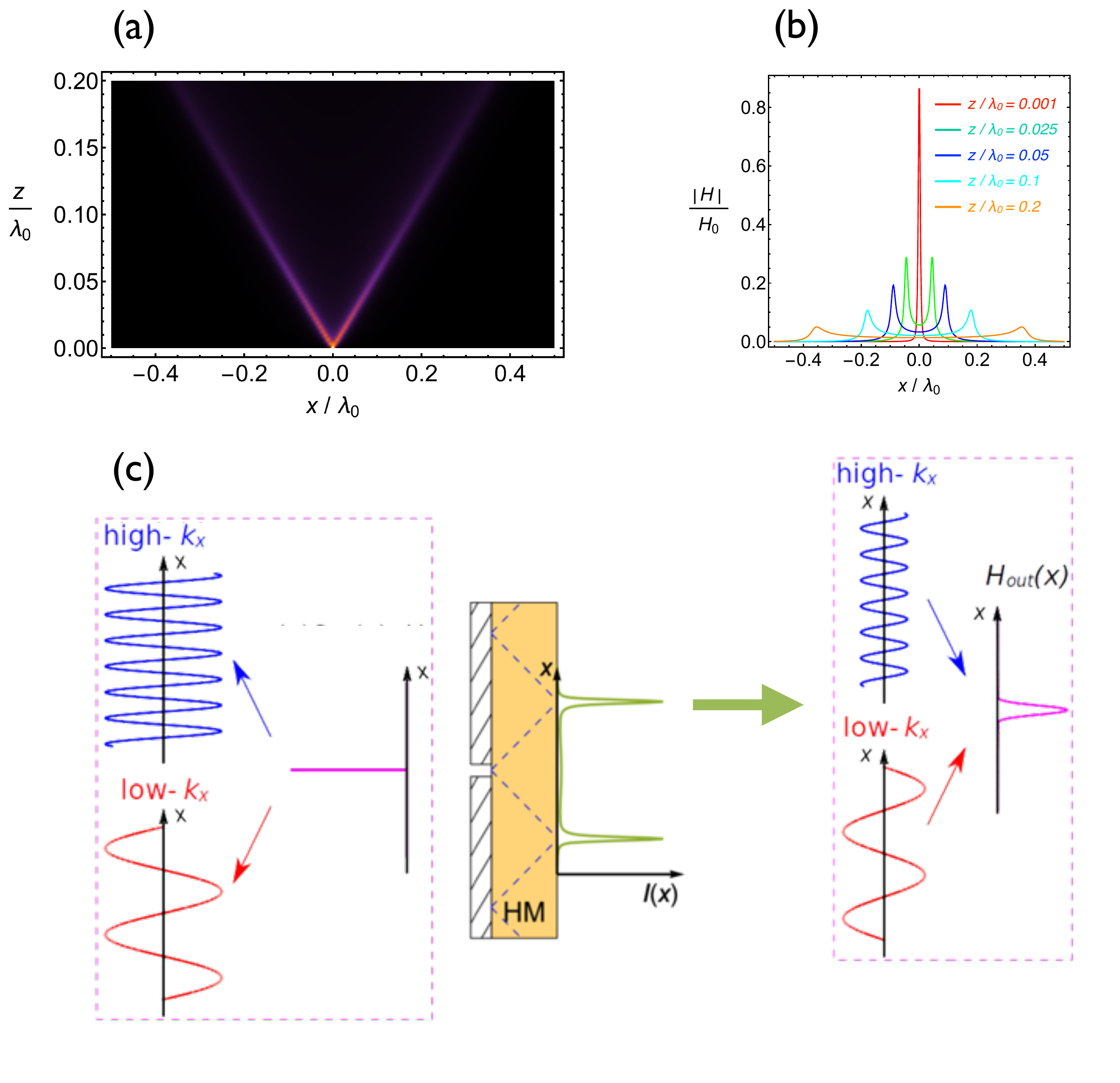}
  \caption{Beam broadening in hyperbolic media. Panel (a): light intensity from a source at the edge of a 
  half-infinite ($z > 0$) hyperbolic material (sapphire). The corresponding free-space wavelength is $\lambda_0 = 20 \ \mu{\rm m}$, and the dielectric permittivity components are $\epsilon_x = \epsilon_y  \simeq 
  7.09+ 0.29 i$, $\epsilon_z \simeq -2.16  + 0.13 i$. The subwavelngth source is formed by a metallic mask with a long slit aligned parallel to the $y$-direction, with the width $a = 20$ nm.  Panel (b): the magnetic field profile at different distances from the surface of the hyperbolic medium, for the same parameters as the panel (a). Note the progressive widening of the beams away from the  source at $x = 0$, $z = 0$. Panel (c):  the schematics illustrating the origin of the degradation of imaging resolution in
 hyperbolic (meta)materials in the presence of material losses. 
}
	\label{fig:HSI}
\end{figure*}

Here we present an alternative  approach to address the problem of optical absorption in metamaterials and plasmonics, that builds upon the pioneering work lead by J.-J. Greffet  \cite{J-J} that demonstrated 
a substantial improvement of the superlens resolution when using time-dependent illumination. We show that,  by
introducing time-variation in the intensity of the incident light with the pattern that is defined by the actual response of the
lossy (meta)material, combined with the time-gating of the transmitted signal, the detrimental  effect of the electromagnetic absorption can be dramatically reduced. The general existence of such ``matching'' solution follows from the fundamental Kramers-Kronig relations, and the ``residual'' absorption is only limited by the signal-to-noise ratio in the system, 
and thus ultimately set by the quantum fluctuations. 

\begin{figure*}[ht!]
  \centering
  \includegraphics[width=0.75\textwidth]{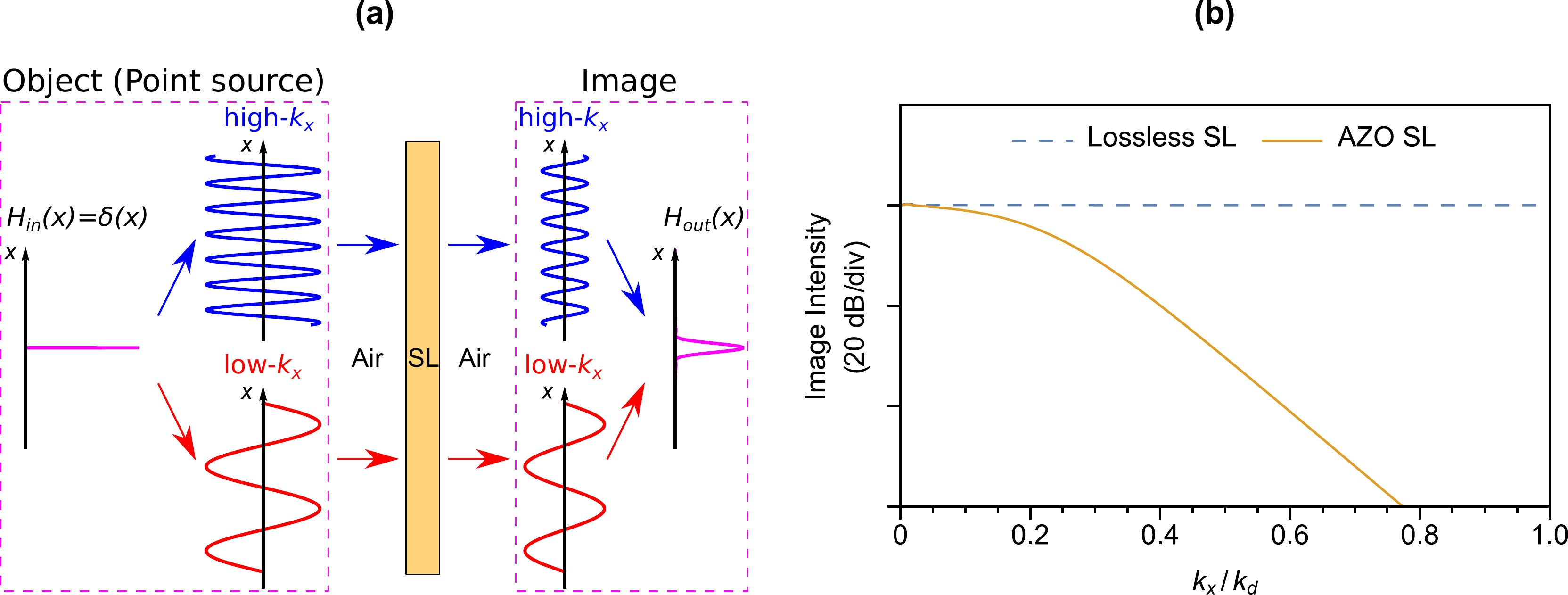}
  \caption{Panel (a): the schematics illustrating the origins of the resolution degradation of the superlens
  in the presence of material losses.
  The lossy near-field superlens (SL) creates a blurred image of the object (CW point source)
  due to uneven attenuation of low-$k_x$ and high-$k_x$ components.
Panel (b):  $k_x$-spectrum at the image plane 
  of a lossless superlens ($\epsilon = -1$, blue dashed line) 
  and a superlens with material loss ($\epsilon \approx -1 + 0.5 i $, orange curve)
  demonstrating the suppression of high-$k_x$ components when illuminated with CW light.
  The data were obtained for a slab (thickness $d=10$~nm) of aluminum-doped zinc oxide (AZO) as a superlens 
  creating an image of the $p$-polarized point light source with the carrier wavelength of 1617~nm 
  corresponding to ${\rm Re}\left[\epsilon\right] = -1$;
  $k_x$ is the transverse wavenumber component and $k_d=\frac{2\pi}{d}$.
}
	\label{fig:SLI}
\end{figure*}

The fundamental impact of material losses in nanophotonics goes substantially beyond simply loosing a significant fraction of the signal power. In plasmonic media and metamaterials with plasmonic components, optical absorption leads to essentially nonuniform attenuation of different transverse wavenumber components. What makes the
matters worse for nanophotonics, is that these high-wavenumber fields that are essential for subwavelength light confinement, experience the
stronger attenuation. This results in a dramatic deterioration of the performance of nanophotonic devices even in the conditions of the relatively small loss. 

\begin{figure}[hb!]
  \centering
  \includegraphics[width=0.5\textwidth]{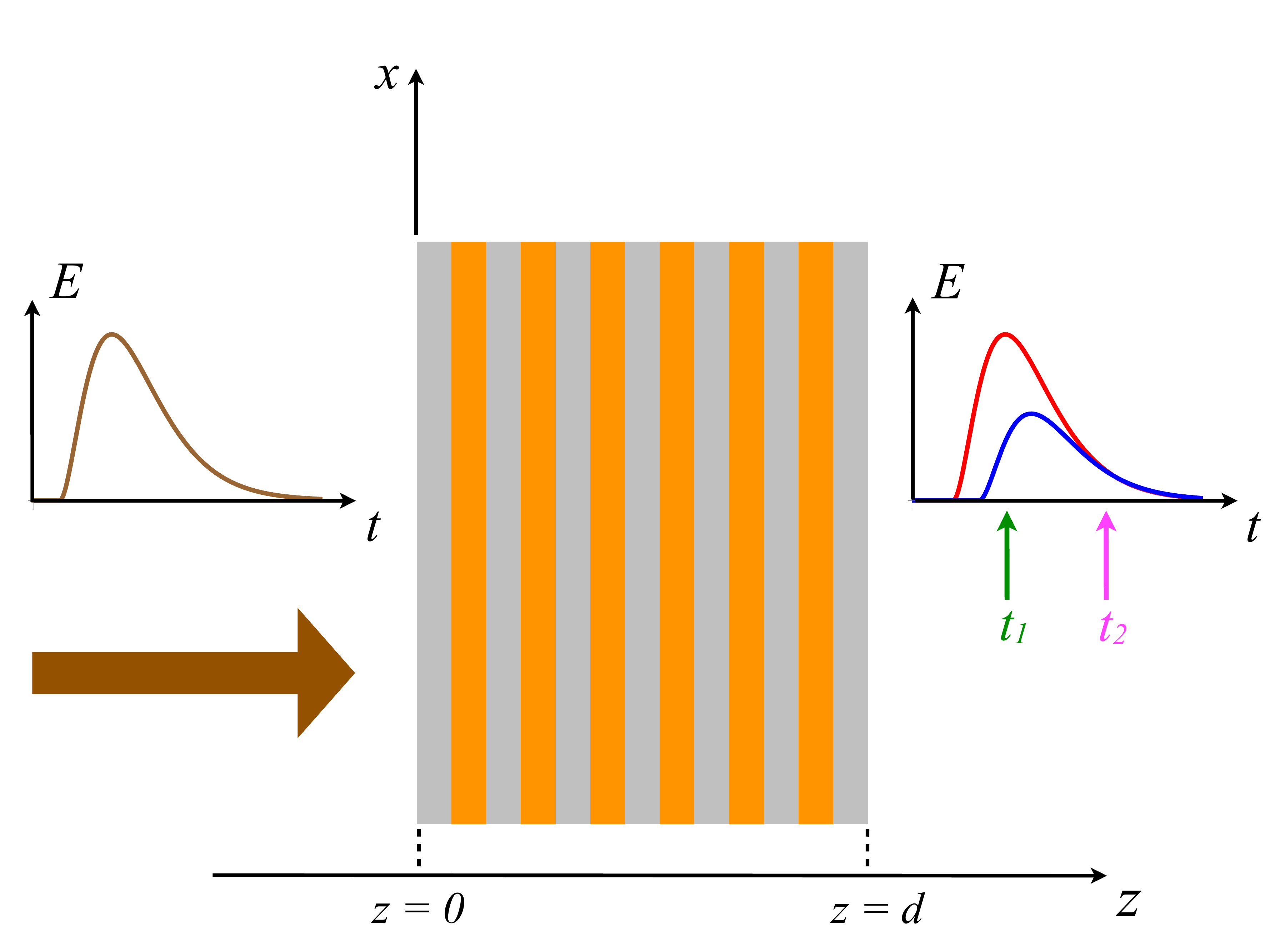}
  \caption{
  The key idea behind the effective loss reduction with pulse shaping. 
  If the illuminating field pulse  profile is appropriately chosen, with the difference in the time delays accumulated over
  the propagation in the metamaterial,
  low-$k_x$ (red curve) and high-$k_x$ (blue curve) 
  components of the image can be power-matched at some point $t_2$ in time.
}
	\label{fig:concept}
\end{figure}

As an example of this behavior, consider light propagation in a hyperbolic (meta)material, where the real parts of the dielectric permittivity components have opposite signs in two orthogonal directions. \cite{podolskiy2005, hyperlens, salandrino2006} In contrast to light in conventional dielectrics where the frequency limits the propagation wavelength and the corresponding light focusing and confinement, the
wavenumbers of propagating fields in hyperbolic media are unrestricted by the frequency. \cite{podolskiy2005} As a result, hyperbolic materials support tightly focused optical beams -- see Fig. \ref{fig:HSI}(a). However, as a function of the propagation distance the relative amplitudes of different wavenumber components in the beam scale (see {\it Methods}) as
\begin{eqnarray}
H_{\ k_\tau \gg \omega / c}  & \propto & \exp\left\{ - \ {\rm Im}\left[\sqrt{-\frac{\epsilon_\tau}{\epsilon_n}}\ \right] \cdot k_\tau \cdot z \right\}, \label{eq:Hk}
\end{eqnarray}
so that the beam looses high-$k$ components at progressively faster rate -- leading to the eventual broadening of the beam in the hyperbolic medium -- see Fig. \ref{fig:HSI}.

Light in the superlens shows a similar behavior -- while the perfect lossless superlens supports all wavenumbers \cite{pendry2000},
introduction of loss effectively suppresses high-$k_x$ components \cite{smith2003,podolskiy2005} -- see Fig.~\ref{fig:SLI}.
With a finite noise that is always present in the imaging system, 
relatively weak high-wavenumber components responsible for spatial resolution 
get lost in the background and the resolution is compromised.\cite{OL-SL}

Note however, that the essential difference in the relative absorption of different wavenumber components, by virtue of 
causality and its mathematical representation in terms of Kramers-Kronig relations, implies a difference in the corresponding
phase velocity and the resulting time delays. As a result, 
if instead of coming from a continuous-wave (CW) source the incident field forms a pulse train, with the proper choice of 
pulse profile, the amplitudes of low- and high-$k_x$ components can be matched at $t > t_2$ -- see Fig.~\ref{fig:concept}).
While detection at an earlier time (e.g. $t_1$) will reveal a  signal with mismatched wavenumber components (and resulting
real-space broadening), time-gating of the output signal near the time $t_2$ will recover the undistorted high-resolution signal.

In order to match the amplitudes of different $k_x$-components at the desired time $t_2$, the required pulse profile (and thus its frequency spectrum) depends on both the material parameters and the dimensions of the metamaterial / plasmonic system.
When the medium satisfies Kramers-Kronig relations, such solution always exists and, as we show in the next section, in most cases is mathematically straightforward.

\begin{figure*}[htb!]
  \centering
  \includegraphics[width=0.9\textwidth]{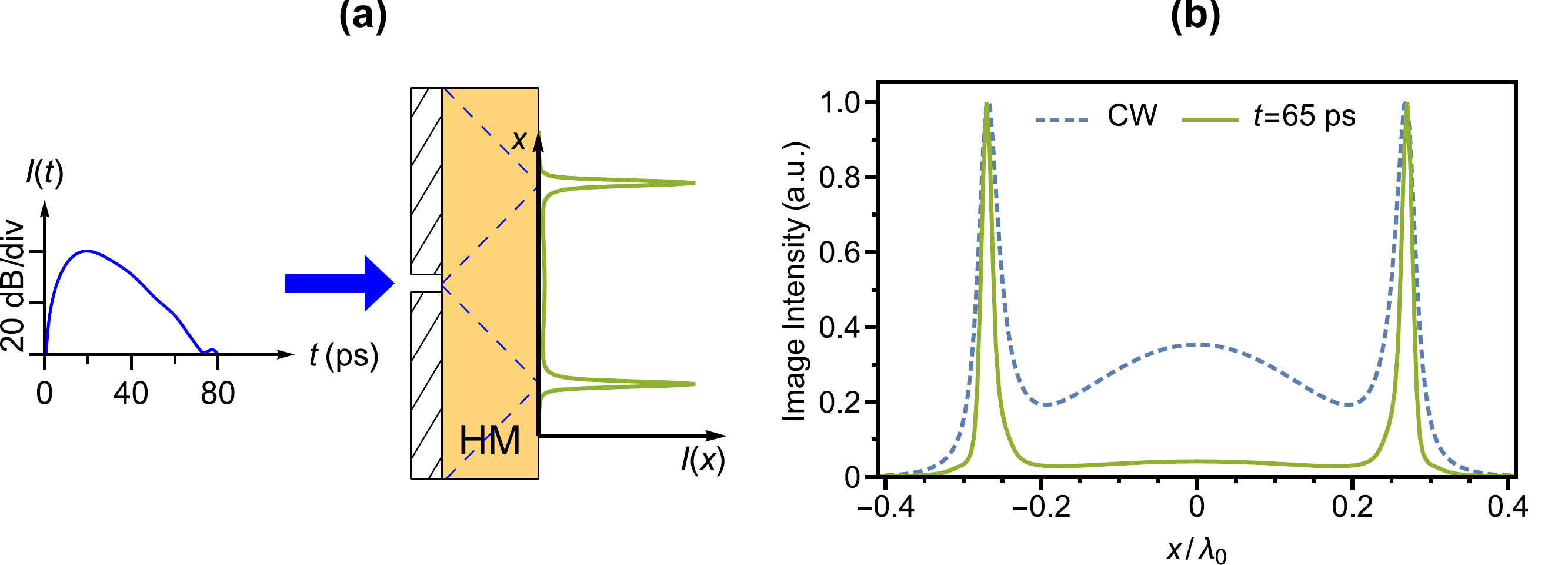}
  \caption{
  Pulse shaping for structured illumination imaging with hyperbolic media (HM).
  Panel (a): the schematics of the imaging system 
  and magnetic field ($B_y$) intensity profile of the shaped light source.
  Panel (b):  the magnetic field intensity profiles
  of the light at the right side surface of the slab
  obtained by illumination with CW light (blue dashed curve) 
  and shaped light (green curve corresponding to the detection time $t^*=65~\rm{ps}$),
  normalized to the unity at their respective maxima.
  The data were obtained for a 3~$\rm{\mu{m}}$ slab of sapphire,
  one side of which is covered with an opaque mask
  with a small aperture in it that effectively creates a point light source on the surface of the slab.
  The mask surface is illuminated with $p$-polarized light at the carrier frequency 
  $f_0 = 15~\mathrm{THz}$ ($\lambda_0 = 20~\mathrm{{\mu}m}$),
  corresponding to the permittivities $\epsilon_e=-2.2 + 0.17i$ and $\epsilon_o=7.1 + 0.29i$
  along the extraordinary and ordinary axes of the sapphire crystal respectively.
  The pulse profile is shaped according to Eq.~(\ref{eq:pulse}), with the  modulation parameter $N=3$
  but the the signal bandwidth strictly limited to the  full spectral width $f_w = 100~\mathrm{GHz}$,
  $\mathrm{FSR} = 7.8~\mathrm{GHz}$.
  The extraordinary axis of the sapphire crystal is normal to the slab surfaces.
}
	\label{fig:fig_sapphire}
\end{figure*}

\section{Results}

In many applications of metamaterials, the observable ${\cal A}_{\rm out} $, in the frequency domain can be related to the original ``input'' ${\cal A}_{\rm in}$ via the linear relation
\begin{eqnarray}
A_{\rm out}\left( \omega\right) & = & \frac{c_0}{\omega - \Omega + i \ \gamma} \
\cdot A_{\rm in}\left(\omega\right)
\label{eq:trangen}
\end{eqnarray}

For example, in the case of subwavelength imaging with a metamaterial superlens, \cite{pendry2000} ${\cal A}_{\rm in}$ represents the original object  pattern, while ${\cal  A}_{\rm out}$ corresponds to the electromagnetic field amplitude measured in the image plane -- see the {\it Methods} section. In this case, while $\gamma$ relates to the loss in the metamaterial, the corresponding parameters $c_0$ and $\Omega$ explicitly depend on the Fourier wavenumber $k_\tau$ (see Eqns, (\ref{eq:cTM}) - (\ref{eq:gTE}) in {\it Methods}), so that the distinct Fourier components of the original pattern are disproportionally represented in the observed image -- with the resulting distortions and loss of resolution. However, in the limit $\gamma \to 0$, at appropriate frequency an exact cancellation reduces this ``transfer function'' to exact unity -- corresponding to the
behavior of a ``perfect lens.'' A finite amount of the effective loss $\gamma$ in Eqn. (\ref{eq:trangen}) will define the actual resolution.

In another example, \cite{HSI} a hyperbolic metamaterial substrate is used to generate subwavelength illumination ``spots'' that can be used for super-resolution imaging within the structured illumination framework, as described in the  {\it Methods} section. In this case, ${\cal A}_{\rm in}$ corresponds to the illumination field amplitude at a given wavelength, that is incident on the hyperbolic  metamaterial substrate -- see Fig. \ref{fig:HSI}, while ${\cal A}_{\rm out}$ represents the field at the other surface of the hyperbolic medium. The spatial variation  of ${\cal A}_{\rm out}$ is defined by the coordinate dependence of $\Omega$ -- see Eqn. (\ref{eq:omega_hsi}) in {\it Methods}. In this example, the Lorenzian structure of the ``transfer function'' of Eqn. (\ref{eq:trangen}) with the spatial coordinate - dependent $\Omega$  corresponds to the (subwavelength) illumination spot  whose central location depends on the illumination frequency. The spatial size of this localized illumination area, set by the effective
material loss parameter $\gamma$ (see Eqns. (\ref{eq:eta_hsi}),(\ref{eq:gamma_hsi}) ),  ultimately defines the imaging resolution of this structured illumination imaging approach that is based on hyperbolic metamaterials.

In the proposed space-time approach, we consider a metamaterial system that can be described by the
general Eqn. (\ref{eq:trangen}), whose ``input'' ${\cal A}_{\rm in}$ (such as e.g. the illumination field in an imaging system) is modulated with a pulse-shaper,
\begin{eqnarray}
{\cal A}_{\rm in}\left(t\right) & = &  {\cal A}\left(t \right) \cdot \exp\left( - i \omega_0 t \right),
\label{eq:Ain}
\end{eqnarray}
in such a wave that the resulting ``input- output'' relation in the time domain
\begin{eqnarray}
{\cal A}_{\rm out}\left( t \right) & = & 
\frac{c_{\rm eff}}{\omega_0 - \Omega + i \ \gamma_{\rm eff}} 
\cdot {\cal A}_{\rm in}\left(t \right)
\label{eq:AAeff}
\end{eqnarray}
has the same mathematical form as (\ref{eq:trangen}) but with a new (possibly time-dependent) effective 
value $\gamma_{\rm eff}$. A substantial reduction of $\gamma_{\rm eff}$ as compared to its original
value $\gamma$, together with appropriate time-gating of the ``output" ${\cal A}_{\rm out}$ is then an equivalent to using a new class of metamaterials with reduced loss.

While somewhat counter-intuitive, this objective can be achieved with a relatively simple pulse shape of the form
\begin{eqnarray}
{\cal A}_N\left(t \right) & = &a_0 \  \theta\left(t\right) \ t^N \exp\left(- \gamma t \right)
\label{eq:pulse},
\end{eqnarray}
where 
$\gamma$ is the material loss parameter as defined in Eqn.~(\ref{eq:trangen}),
$N \geq 0$  does not need to be an integer, 
and $\theta\left( t\right)$ is the Heaviside function (zero for a negative argument, and unity otherwise). Note that the frequency spectrum of (\ref{eq:pulse}) is a simple power law, so that its actual implementation with a practical pulse-shaper should be straightforward. 

\begin{figure*}[htb!]
  \centering
  \includegraphics[width=0.9\textwidth]{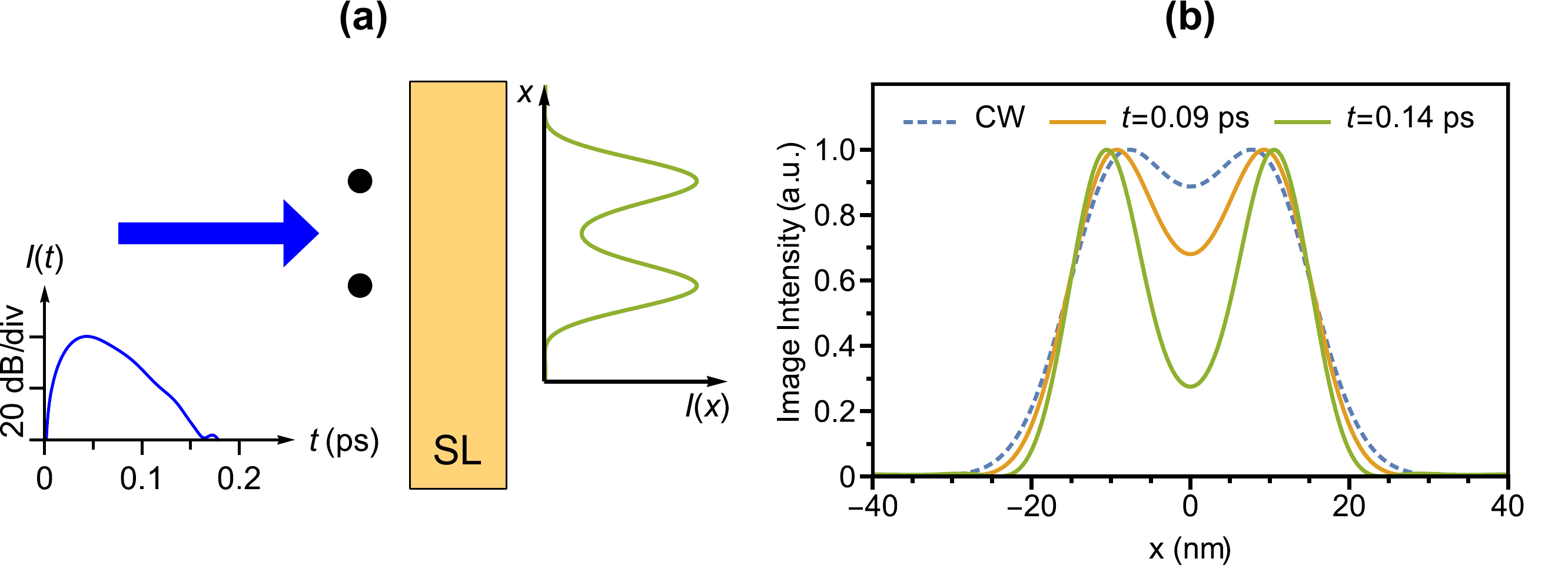}
  \caption{
  Pulse shaping for imaging for a superlens (SL).
Panel (a): the schematics of the imaging system 
  and magnetic field ($H_y$) intensity profile of the shaped light source.
Panel (b): the magnetic field intensity 
  of the images obtained by illuminating the target 
  with CW light (blue dashed curve) 
  and with the pulse train ``tuned'' to the material response (orange and green curves corresponding to the detection times 
  $t_1=0.09~\rm{ps}$ and $t_2=0.14~\rm{ps}$ respectively),
  normalized to the unity at their respective maxima.
    The imaging data were calculated for a 10~nm slab of aluminum-doped zinc oxide (AZO) 
    as a near-field superlens 
  creating an image of two point objects located 20~nm from each other. 
  The SL is illuminated with $p$-polarized light source at the carrier frequency 
  $f_0 = 185~\mathrm{THz}$ ($\lambda_0 = 1617~\mathrm{{n}m}$),
  corresponding to ${\rm Re}\left[\epsilon\right] = -1$.
  The pulse modulation parameter $N=3$.
  In the simulation, the superlens is illuminated with a period pulse train, 
with free spectral range $\mathrm{FSR} = 3.5~\mathrm{THz}$.
  The pulse spectrum is limited to 
  full bandwidth $f_w = 44~\mathrm{THz}$.
}
	\label{fig:SL}
\end{figure*}

With (\ref{eq:pulse}), for ${\cal A}_{\rm out}$ in the time domain we find
\begin{eqnarray}
{\cal A}_{\rm out}\left( t \right) & = & 
-i \ c_0 \  t \cdot {\cal F}_N\left[\left(\omega_0 - \Omega\right)t\right]
\cdot {\cal A}_{\rm in}\left(t \right),
\label{eq:AF}
\end{eqnarray}
where
\begin{eqnarray}
{\cal F}_N\left(x\right) & = & \frac{N!}{\left(i x\right)^{N+1}} \cdot \left(\exp\left(i x \right) - \sum_{k = 0}^{N} \frac{\left(i x\right)^k}{k!} \right) \\
& = & \frac{\exp\left(i x \right)}{\left(i x\right)^{N+1}}\  \Big( N! - \Gamma\left(N+1,i x\right) \Big),
\label{eq:FNx} 
\end{eqnarray}
and $\Gamma\left(n,x\right)$ is the incomplete Gamma-function.

Using the Pad\'e approximation \cite{Pade} for the function ${\cal F}\left(x\right)$ (see {\it Methods}), we obtain
\begin{eqnarray}
{\cal F}_N\left(x\right) & \simeq & \frac{1}{N + 1 - i x},
\label{eq:Pade}
\end{eqnarray}
which together with (\ref{eq:AF}) yields the desired Eqn. (\ref{eq:AAeff}), with the effective parameters
\begin{eqnarray}
c_{\rm eff} & = & c_0, \\
\gamma_{\rm eff} & = &  \frac{N+1}{t}. \label{eq:g_eff}
\end{eqnarray}

Time-gating the output ${\cal A}_{\rm out}\left( t \right)$ at $t > 1 /  \left( N + 1 \right)\gamma$ will then reduce the effective width $\gamma_{\rm eff} $ beyond the original value $\gamma$ that was defined by the actual material parameters. Furthermore, as long as the photonic system is adequately described by  the classical Maxwell equations (so that the quantum noise can be neglected), there is no limit on 
the degree of the reduction of the ``effective'' loss. As the fundamental level, the proposed space-time
 approach therefore solves the loss problem that plagued the field of metamaterials for the last decade.
 
However, with the pulse profile (\ref{eq:pulse}), time-gating the output signal at $t \gg 1/\gamma$ implies  operating at progressively lower powers. As a result, there is a practical limit to the ``loss mitigation'' in the proposed space-time approach that is defined by the actual signal-to-noise power ratio ${\rm SNR}$  in the system. 
 
From Eqns. (\ref{eq:pulse}) and (\ref{eq:g_eff}), we obtain
\begin{eqnarray}
\gamma_{\rm eff} & = & \gamma \ \frac{N + 1}{ N + \frac{1}{2}  \log\frac{P_{\rm peak}}{P\left(t\right)} },
\end{eqnarray}
where $P_{\rm peak}$ is the peak power of the pulse -- which defines the limit to the effective loss
reduction  in the proposed space-time approach 
\begin{eqnarray}
\gamma_{\rm min} & = & \frac{2  \gamma }{ \log{\rm SNR}}.
\end{eqnarray}

For super-resolution imaging, this result can also be interpreted in terms of the effective loss, e.g.
the imaginary part of the dielectric permittivity of the (meta)material that would allow the same
resolution power with CW illumination, as our approach:
\begin{eqnarray}
\epsilon_{\rm eff}'' \simeq \epsilon'' \ \frac{N + 1}{ N + \frac{1}{2}  \log\frac{P_{\rm peak}}{P} },
\label{eq:eps_eff}
\end{eqnarray}
where $ \epsilon''$ corresponds to the actual value of the permittivity. In case of a large signal-to-noise ratio,
our approach therefore offers an intriguing alternative to the search of new and better metamaterials and 
plasmonic media -- instead of trying to reduce the actual material absorption, the same result can be 
achieved with the temporal modulation of the incident light that is ``tuned'' to the response of the existing media (note that 
the pulse parameter $\gamma$ in Eqn. (\ref{eq:pulse}) is defined by the material response in Eqn. (\ref{eq:trangen}) ).

\section{Discussion}

Based on the general linear response formulation of Eqn. (\ref{eq:trangen}), our 
proposed space-time approach can be applied to a broad range of nanophonic  systems that involve
metamaterial or plasmonic elements. In the first example, we apply it to  hyper-structured illumination imaging, \cite{HSI} using 
sapphire in its hyperbolic band near 20~$\rm{\mu{m}}$ -- see Fig. \ref{fig:fig_sapphire}. In this simulation, to adequately represent a practical experimental
setup, we furthermore assume that the pulse-shaper is only operating in a finite bandwidth window, with all the frequency 
components of the band-limited signal within its range. With the pulse-shaping of the incident field 
and the time-gating of the transmitted light at $t^*=65~\rm{ps}$, we achieve
the same field localization as if the imaginary parts of the dielectric permittivity at both ordinary and extraordinary axes 
were effectively reduced by approximately 50\% 
-- from ${\rm Im}\left[\epsilon\right] \approx 0.29$ to ${\rm Im}\left[\epsilon_{\rm eff} \right] \approx 0.13$ in the ordinary direction,
and from ${\rm Im}\left[\epsilon\right] \approx 0.17$ to ${\rm Im}\left[\epsilon_{\rm eff} \right] \approx 0.075$ in the extraordinary one.

As the other example, we consider the application of the proposed space-time approach to
super-resolution imaging with 
the near-field superlens based on aluminum-doped zinc oxide (AZO), operating at
the telecommunication wavelength of $\mathrm{1617~nm}$ -- see Fig.~\ref{fig:SL} . 
While this superlens is originally
unable to resolve two point objects at $20$ nm spacing due to the material loss of the AZO, 
using the pulse-shaped illumination 
with full spectral width $f_w = 44~\mathrm{THz}$
and free spectral range $\mathrm{FSR} = 3.5~\mathrm{THz}$
immediately solves the problem -- see
Fig.~\ref{fig:SL}(b). These results are consistent with the original work of \cite{J-J} which
demonstrated an improvement of the superlens focusing properties for time-dependent 
illumination.

In the practical implementation of our proposed space-time metamaterial approach, 
the key challenges lie in the actual realization  of ultrafast pulse shaping and ultrafast temporal 
measurement and characterization.
With active research in these areas  over the last two decades, the necessary tools for ultrafast pulse shaping 
and arbitrary waveform generation on picosecond scale and beyond are already available. \cite{weiner2011,ferdous2011}
However, the time-gating at the relatively low signal powers that is essential for our approach, remains a challenging task. 
For example, in the case of the hyper-structured illumination with sapphire, one needs ultrafast measurement of optical pulses
in the far infrared region. While the standard  GaAs and InGaAs detectors are primarily limited to the near infrared spectral range,
there have been proposed several mechanisms to achieve ultrafast detection in the far infrared,
including rectification in field-effect transistors \cite{preu2013}, Schottky diodes \cite{semenov2010}, or superlattice detectors \cite{klappenberger2001}. For picosecond and sub-picosecond optical pulse measurement that would be required
for implementation of the suggested approach for the superlens imaging with AZO ($\lambda_0 = 1617~\mathrm{{n}m}$) 
 there are also a number of techniques that are already available, such as temporal magnification (or ``time-lens'') \cite{bennett1994,bennett1999,foster2008,pasquazi2010}
and time-to-frequency conversion. \cite{kauffman1994,azana2004} We therefore conclude that, 
our space-time metamaterial  approach can be implemented entirely within the constraints of the 
currently available  technology.

\section{Methods}

In this section, we provide the theoretical background for Eqn. (\ref{eq:trangen}) that describes electromagnetic field in a range of nanophotonic systems -- from hyperbolic media to plasmonic systems and negative index metamaterials.

\subsection{Light in hyperbolic media}

With the opposite signs of the dielectric permittivity components in two orthogonal directions, hyperbolic media can support TM-polarized electromagnetic fields with the wavenumbers that are only limited by the size of the unit cell of the material. For a uniaxial hyperbolic medium with the permittivities $\epsilon_x = \epsilon_y$ and $\epsilon_z$, for the relation between the wavevector ${\bf k}$ and the frequency $\omega$ we find \cite{hyperlens} 
\begin{eqnarray}
\frac{k_x^2 + k_y^2}{\epsilon_z} + \frac{k_z^2}{\epsilon_x} & = & \frac{\omega^2}{c^2}, \label{eq:dispersion}
\end{eqnarray}
For a thin (subwavelength)  illumination slit at the surface of the hyperbolic medium, corresponding to the configuration in Fig. 1, the TM-polarized  magnetic field in  the hyperbolic material, 
\begin{eqnarray}
{\bf H}\left({\bf r}, t\right)  =  \int d\omega  \ H_\omega\left(x,z\right) \  \exp\left(- i \omega t\right) \hat{\bf y}
\label{eq:B_hm}
\end{eqnarray}
where its time-harmonic amplitudes $H_\omega\left(x, z\right)$ can be expressed as
\begin{eqnarray}
H_\omega\left(x, z\right) & = & \frac{H_\omega}{2 \pi} \int_{- \infty}^\infty dk_x \exp\left(
i k_x x \right) 
\nonumber \\
& \times & \exp\left( i \sqrt{\epsilon_x \left(\omega/c\right)^2 - \left(\epsilon_x/\epsilon_z\right) k_x^2} \ s \ z
\right)
\label{eq:B_amplitude}
\end{eqnarray}
where 
\begin{eqnarray}
s & = & {\rm sign} \left[ \  {\rm Im} \sqrt{ - \frac{\epsilon_x}{\epsilon_z} } \  \right]
\end{eqnarray}
The expression (\ref{eq:B_amplitude}) is exact for a natural hyperbolic medium, and is an accurate approximation for a hyperbolic metamaterial when its unit cell size is on the same order or smaller than 
the width of the illumination slit.

Within the absorption distance from a narrow (compared to the free-space wavelength) illumination field, the integral (\ref{eq:B_amplitude}) is dominated by the wavenumbers $k_x \gg \omega/c$. The 
$z$-component of the wavevector can then be approximated by
\begin{eqnarray}
k_z \equiv \sqrt{\epsilon_x \left(\omega/c\right)^2 - \left(\epsilon_x/\epsilon_z\right) k_x^2}
\simeq \sqrt{ - \epsilon_x/\epsilon_z} \ \left| k_x\right|,
\end{eqnarray}
which corresponds to the linear asymptotic behavior of the dispersion relation (\ref{eq:dispersion}).
The integral (\ref{eq:B_amplitude}) can then be calculated analytically, which yields
\begin{eqnarray}
H_\omega\left(x, z\right) & = &
\frac{H_\omega}{2\pi i } \ 
 \sum_\pm \frac{1}{-  s \sqrt{- \frac{\epsilon_x}{\epsilon_z}} \ z \pm  x}
 \label{eq:B_asym}
\end{eqnarray}

\begin{figure}[ht!]
  \centering
  \includegraphics[width=0.45\textwidth]{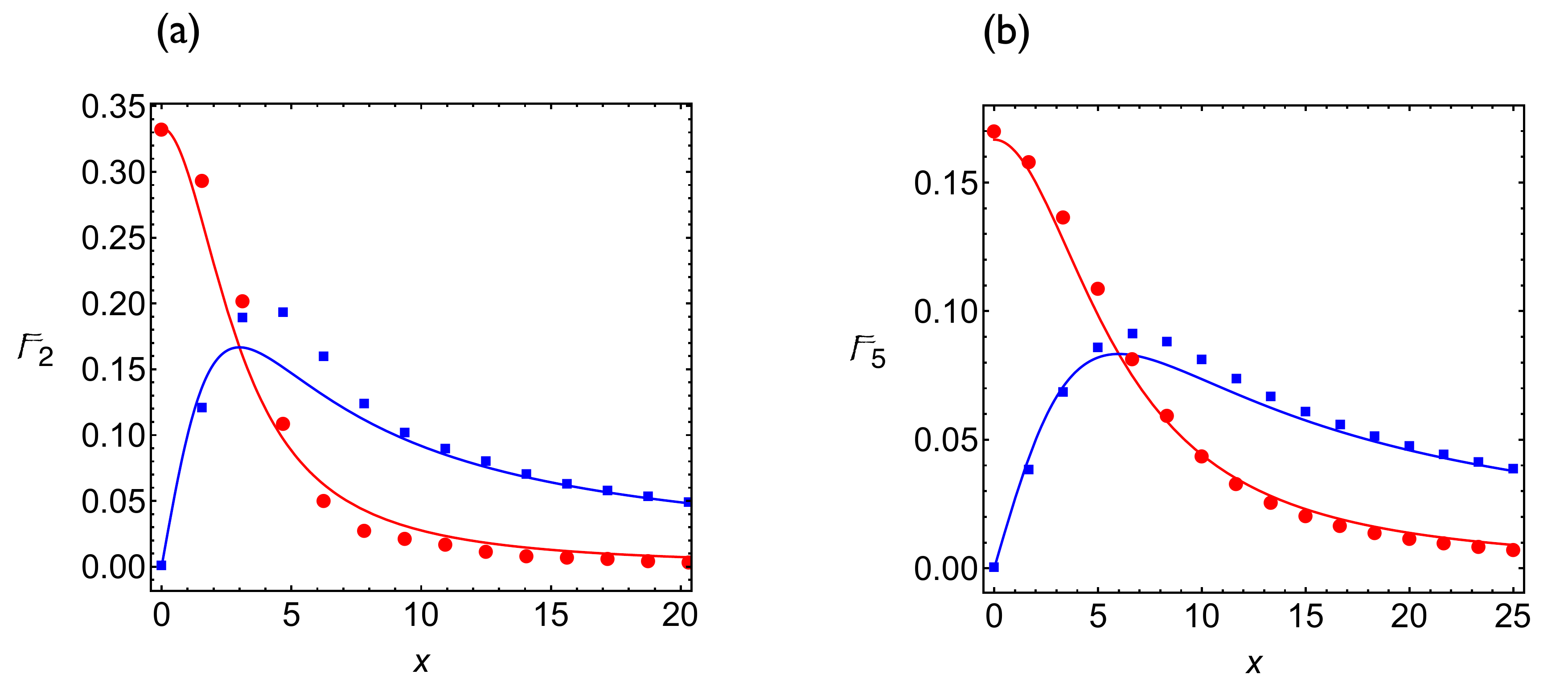}
  \caption{
 The exact values of ${\cal F}_N$ (symbols) and the corresponding Pad\'e approximation (solid lines), for $N = 2$ (a) and $N = 5$
 (b). Red color shows the values of the real parts of ${\cal F}_N$ and ${\cal F}_N^{ \ \cal P}$, while blue represents the corresponding imaginary components.
}
	\label{fig:Pade}
\end{figure}

If the frequency bandwidth of the ``signal'' $H\left(t\right)$, centered around the frequency $\omega_c$,  is much smaller than the characteristic scale corresponding to a substantial variation of any of the components of the permittivity tensor of the
hyperbolic medium, we can approximate 
\begin{eqnarray}
s \sqrt{- \frac{\epsilon_x}{\epsilon_z}} \simeq \eta_0' + \eta_1' \left(\omega - \omega_c\right) + i \eta'',
\label{eq:eta_hsi}
\end{eqnarray}
which reduces (\ref{eq:B_asym}) to a special case of our general expression (\ref{eq:trangen}),
\begin{eqnarray}
H_\omega\left(x, z\right) & \simeq & H_\omega \sum_\pm \frac{c_h}{\omega - \Omega_\pm\left(x,z\right) + i \gamma_h},
\end{eqnarray}
where
\begin{eqnarray}
c_h & = & \frac{i}{2 \pi \ z \ \eta_1'}, \\
\gamma_h & = & \frac{\eta_h''}{\eta_1'}, \label{eq:gamma_hsi}
\end{eqnarray}
and
\begin{eqnarray}
 \Omega_\pm\left(x,z\right)  & = & \omega_c - \frac{\eta_0'}{\eta_1'} \pm \frac{x}{\eta_1' z}.
 \label{eq:omega_hsi}
\end{eqnarray}

\subsection{Electromagnetic field in the superlens}

In its simple realization, \cite{pendry2000} the superlens is essentially a parallel slab of a metamaterial with simultaneously negative values of the (isotropic) dielectric permittivity $\epsilon$ and magnetic permeability $\mu$, that ``match'' the corresponding parameters of the surrounding medium $\epsilon_0$, $\mu_0$:
\begin{eqnarray}
\epsilon & = & - \ \epsilon_0,  \label{eq:eps_res} \\
\mu & = & - \ \mu_0. \label{eq:mu_res}
\end{eqnarray}
A perfect resonance of this kind is however unattainable, due to inevitable  finite amount of loss in the metamaterial, leading to nonzero imaginary parts of $\epsilon$ and $\mu$. 

The planar superlens produces a (nearly) perfect image when the separation between the ``object''' and ``image'' plane is equal to twice the thickness of the superlens $d$, with the corresponding transmission coefficient  as a function of the in-plane wavenumber $k_\tau$, equal to 
\begin{eqnarray}
T_\omega^{\rm TM}\left(k_\tau\right) 
& = & \frac{4 \  \exp\left[\left(\kappa - \kappa_0 \right) d\right] }{\left(2 +\frac{\kappa}{{\epsilon}  \kappa_0} + \frac{{\epsilon}  \kappa_0}{\kappa}\right) \ \exp\left(2 \kappa d\right) +\left(2 - \frac{\kappa}{\epsilon  \kappa_0} - \frac{ {\epsilon} \kappa_0}{\kappa}\right) } \ \ \ \ \ \\
& = & \frac{2 \epsilon \kappa_0 \kappa}{\epsilon \kappa_0 - \kappa }  \exp\left[\left(\kappa - \kappa_0 \right) d\right]  \nonumber \\
& \times & 
\sum_\pm \frac{1}{\left(\kappa - \epsilon \kappa_0\right) \pm \left(\kappa + \epsilon \kappa_0\right)\ \exp\left( \kappa d\right)},
\label{eq:T_TM_exact}
\end{eqnarray}
for TM-polarized fields, and
\begin{eqnarray}
T_\omega^{\rm TE}\left(k_\tau\right) 
& = & \frac{4 \  \exp\left[\left(\kappa - \kappa_0 \right) d\right] }{\left(2 +\frac{\kappa}{{\mu}  \kappa_0} + \frac{{\mu}  \kappa_0}{\kappa}\right) \ \exp\left(2 \kappa d\right) +\left(2 - \frac{\kappa}{{\mu}  \kappa_0} - \frac{ {\mu} \kappa_0}{\kappa}\right) } \ \ \ \ \  \\
& = & \frac{2 \mu \kappa_0 \kappa}{\mu \kappa_0 - \kappa }  \exp\left[\left(\kappa - \kappa_0 \right) d\right]  \nonumber \\
& \times & 
\sum_\pm \frac{1}{\left(\kappa - \mu \kappa_0\right) \pm \left(\kappa + \mu \kappa_0\right)\ \exp\left( \kappa d\right)}, 
\label{eq:T_TE_exact}
\end{eqnarray}
for the TE-polarization, where
\begin{eqnarray}
\kappa_0  \equiv \sqrt{k_\tau^2 - \omega^2 / c^2},
\end{eqnarray}
and
\begin{eqnarray}
\kappa  \equiv \sqrt{k_\tau^2 - \epsilon \mu \   \omega^2 / c^2}.
\end{eqnarray}
In the limit $\epsilon/\epsilon_0 , \ \mu/\mu_0  \to -1$, we find $T_\omega\left(k_\tau\right) \to 1$ for any $k_\tau$ -- which implies the formation of a perfect image.

Note that the concept of super-resolution relies on the accurate representation of high-$k$ Fourier components of the object pattern, $k_\tau \gg \omega/c$. In this limit, for a small loss $\epsilon'' \ll \left|\epsilon\right|$, $ \mu'' \ll \left| \mu\right|$ and the signal bandwidth smaller than the frequency scale corresponding to a substantial variation of $\epsilon$ and $\mu$,  we find 
\begin{eqnarray}
T_\omega^{\rm TM}\left(k_\tau\right) & \simeq &  \sum_\pm \frac{c_\pm^{\rm TM}}{\omega - \Omega_\pm^{\rm TM}\left(k_\tau\right) + i \gamma^{\rm TM}},
\label{eq:T_TM}
\end{eqnarray}
and
\begin{eqnarray}
T_\omega^{\rm TE}\left(k_\tau\right) & \simeq &  \sum_\pm \frac{c_\pm^{\rm TE}}{\omega - \Omega_\pm^{\rm TE}\left(k_\tau\right) + i \gamma^{\rm TE}},
\end{eqnarray}
where
\begin{eqnarray}
c_\pm^{\rm TM} & = & \pm \left. \frac{ \  \exp\left( - \left| k_\tau \right|  d\right)}{{d\epsilon'}/{d\omega}} \ \right|_{\ \omega \  = \ \omega_c}  , \label{eq:cTM} \\ 
 \Omega_\pm^{\rm TM}\left(k_\tau\right) & = & \omega_c \mp 
\left. 
 \frac{2  \exp\left(- \left| k_\tau\right|  d\right)}{{d\epsilon'}/{d\omega}}  \ \right|_{\ \omega \  = \ \omega_c} ,  \\
\gamma^{\rm TM} & = & \left. \frac{\epsilon''}{{d\epsilon'}/{d\omega}} \ \right|_{\ \omega \  = \ \omega_c} , 
\end{eqnarray}
and
\begin{eqnarray}
c_\pm^{\rm TE} & = & \pm \left. \frac{ \  \exp\left( - \left| k_\tau \right|  d\right)}{{d\mu'}/{d\omega}} \ \right|_{\ \omega \  = \ \omega_c}  , \label{eq:cTE} \\ 
 \Omega_\pm^{\rm TE}\left(k_\tau\right) & = & \omega_c \mp 
\left. 
 \frac{2  \exp\left(- \left| k_\tau\right|  d\right)}{{d\mu'}/{d\omega}}  \ \right|_{\ \omega \  = \ \omega_c} ,  \\
\gamma^{\rm TE} & = & \left. \frac{\mu''}{{d\mu'}/{d\omega}} \ \right|_{\ \omega \  = \ \omega_c} . \label{eq:gTE}
\end{eqnarray}
As a result, the transmission function of a superlens can also be represented by our general expression (\ref{eq:trangen}).

Note that, for super-resolution imagining in the near field, the desired functionality can be also achieved by the so called ``poor man's superlens,'' that is essentially a metallic layer used for TM-polarized light near the plasmon resonance frequency that corresponds to Eqn. (\ref{eq:eps_res}). The corresponding transmission coefficient, also given by Eqn. (\ref{eq:T_TM_exact}) but with the value of $\kappa$ taken at $\mu = 1$, 
in the subwavelength resolution resolution limit $k_\tau \gg \omega/c$ then also reduces to Eqn. (\ref{eq:T_TM}) that can be treated as a special case of Eqn. (\ref{eq:trangen}).

\subsection{Pad\'e Approximation}

For a given function $f\left(x\right)$ in a specified interval $(a, b)$, the  Pad\'e approximant is the rational function whose power series expansions near $x = a$ and $x= b$ agree with the corresponding power series expansions of $f\left(x\right)$ to the highest possible order. \cite{Pade}  In our case, we seek to approximate ${\cal F}_N\left(x\right)$ in the entire range $0 \leq x < \infty$, so that with Pad\'e approximant of the first order,
\begin{eqnarray}
{\cal F}_N^{\cal \ P}\left(x\right) & = & \frac{A_N}{B_N + C_N x},
\end{eqnarray} 
we impose the conditions
\begin{eqnarray}
\frac{A_N}{B_N}  & = & {\cal F}_N\left( 0\right), \\
\frac{A_N}{C_N} \frac{1}{x}  & = &  \frac{1}{x} \cdot \lim_{x \to \infty} \left[ x \ {\cal F}_N\left( x\right) \right],
\end{eqnarray}
which yields 
\begin{equation}
\frac{A_N}{B_N}  =  \frac{1}{N},\ \ \ \ 
\frac{A_N}{C_N}  =  i,
\end{equation}
and therefore
\begin{eqnarray}
{\cal F}_N^{\cal \ P}\left(x\right) & = & \frac{1}{N + 1 - i x},
\label{eq:Pade1}
\end{eqnarray}
which immediately leads to Eqn. (\ref{eq:Pade}).

In Fig. \ref{fig:Pade} we compare the exact values of ${\cal F}_N$ for $N = 2$ (a) and $N = 5$ (b) with the corresponding Pad\'e
approximation of Eqn. (\ref{eq:Pade1}). Note that our first order  Pad\'e approximant (\ref{eq:Pade1})  offers an accurate representation of the
exact function ${\cal F}(x)$ in the entire range $0 \leq x < \infty$. 

\section{Acknowledgements}
This work was partially
supported by Gordon and Betty Moore Foundation and NSF Center for
Photonic and Multiscale Nanomaterials (C-PHOM).
The authors thank Prof. A. M. Weiner for the helpful discussions and comments on the article.

\end{document}